\documentclass{PoS}

\usepackage{amsfonts,amssymb,amsmath} %amsthm
\usepackage{cite}

\title{Renormalisation of the energy-momentum tensor in scalar field theory using the Wilson flow}

\ShortTitle{Energy-momentum tensor in scalar field theory from the Wilson flow}

\newcommand{\edimburgo}{Higgs Centre for Theoretical Physics, School of Physics and Astronomy,
  University of Edinburgh, Edinburgh EH9 3FD, UK}
\newcommand{\plymouth}{Centre for Mathematical Sciences, Plymouth University, Plymouth PL4 8AA, UK}

\author{Francesco Capponi, Antonio Rago
	 \\
	 \plymouth\\
	 E-mail: \email{francesco.capponi@plymouth.ac.uk, antonio.rago@plymouth.ac.uk}}
\author{Luigi Del Debbio, \speaker{Susanne Ehret}, Roberto Pellegrini
	 \\
	 \edimburgo\\
	 E-mail: \email{luigi.del.debbio@ed.ac.uk, susanne.ehret@ed.ac.uk, rpellegr@staffmail.ed.ac.uk}}

\abstract{A non-perturbative renormalisation prescription for the energy-momentum tensor, based on space-time symmetries along the Wilson flow, has been proposed recently in the context of 4-dimensional gauge theories. We extend this construction to the case of a scalar field theory, and investigate its numerical feasibility by studying Ward identities in 3-dimensional scalar field theory. After introducing the Wilson flow for the scalar field theory we discuss its renormalisation properties and the determination of the renormalisation constants for the energy-momentum tensor.}

\FullConference{The 33rd International Symposium on Lattice Field Theory\\
                 14 -18 July  2015\\
                 Kobe International Conference Center, Kobe, Japan}

\begin{document}

\newcommand{\tr}{\mbox{tr}} 
\newcommand{\e}{\mbox{e}}

\section{Introduction}

The Wilson flow is a promising tool to study strongly coupled theories on the lattice. Recent studies of the renormalisation of the coupling and composite operators on the lattice involving the Wilson flow prove the success of the method, see e.g. refs. \cite{Lin:2015zpa,Ramos:2015baa,Ramos:2015dla} and references therein. The application of the Wilson flow for the renormalisation of the energy-momentum tensor has been proposed in ref. \cite{DelDebbio:2013zaa}. For an alternative approach to determine the energy-momentum tensor see e.g. refs. \cite{Suzuki:2013gza,Giusti:2015daa}. We further the investigation of the Wilson flow by studying its use for the non-perturbative computation of the energy-momentum tensor in 3-dimensional scalar field theory. Choosing a single component scalar field theory in only 3 dimensions gives us the opportunity to gain a deeper understanding on a theoretical and numerical level as it has the advantage of low computational costs while maintaining high statistics.  

Scalar field theory in 3 dimensions with negative mass squared exhibits an infrared fixed point and can thus serve as a toy model for the research of more advanced theories with an infrared fixed point. Studying the energy-momentum tensor in this context will enable us to investigate the scaling behaviour of such theories, as the trace of the energy-momentum tensor $T_{\mu\mu}$ is related to the beta-function,
\begin{equation}
  \hspace{-1.5cm} \langle \int d^Dx\, T_{\mu\mu}\, \phi(x_1) ... \phi(x_n) \rangle = -\left(\sum_k \beta_k \frac{\partial}{\partial g_k} + n(\gamma_{\phi} + d_{\phi}) \right) \langle \phi(x_1) ... \phi(x_n) \rangle,
\end{equation}
where $\beta_k$ are the beta functions, $g_k$ are the couplings of the theory, $\gamma_{\phi}=-\frac{1}{2Z_{\phi}}{\scriptstyle \mu}\frac{\text{d}}{\text{d}\mu} {\scriptstyle Z_{\phi}}$, $Z_{\phi}$ is the field renormalisation factor, and $d_{\phi}$ is the dimension of field $\phi$. This relation can be found by using the Callan-Symanzik equation and the dilatation Ward identity.

\begin{figure}
\centering
  \includegraphics[width=6.5cm,trim={0cm 0cm 0cm 1cm},clip]{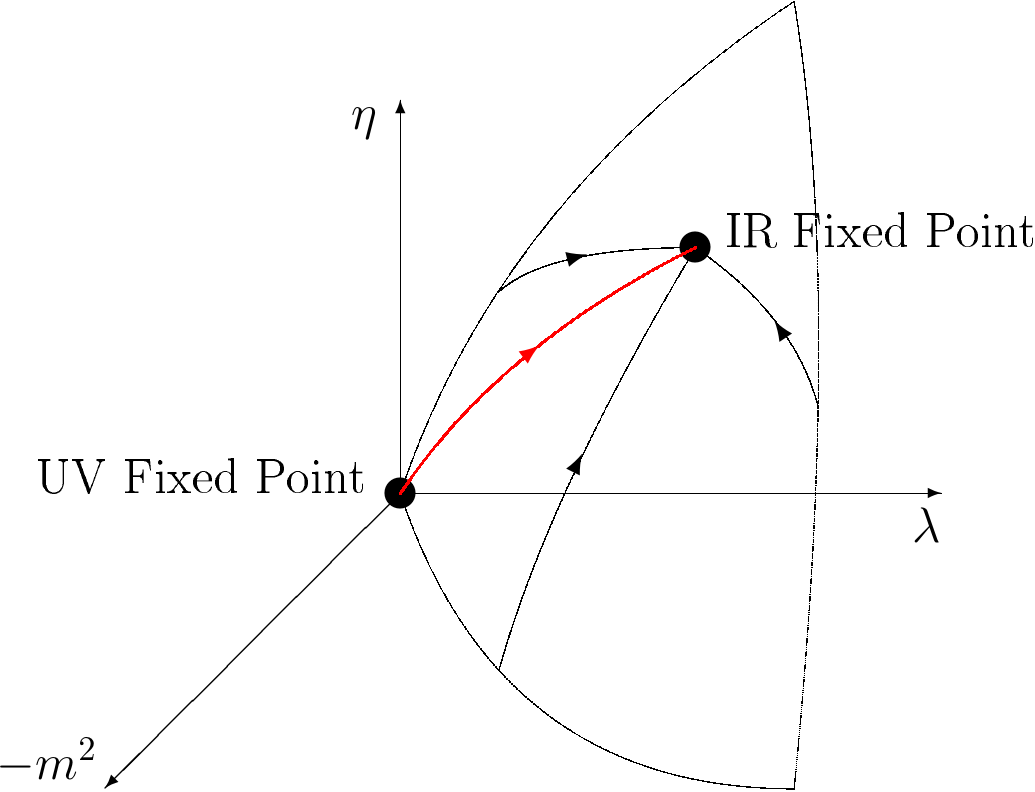}
\label{fig:RGflowdiag}
\caption{Sketch of coupling space and RG flow of $\phi^6$ theory.}
\end{figure}

There are two relevant operators and one marginal operator near the Gaussian fixed point in 3 dimensions, $\phi^2$, $\phi^4$ and $\phi^6$ with couplings $m^2$, $\lambda$ and $\eta$, respectively.
The coupling space and renormalisation group flow of the theory are sketched in fig. \ref{fig:RGflowdiag}. There are two fixed points, a Gaussian or ultraviolet fixed point, and a Wilson-Fischer or infrared fixed point. The surface indicated is the critical surface where the dimensionful renormalised couplings $m_R$ and $\lambda_R$ are zero, and physics is constant. On this surface the renormalisation group flow is directed toward the infrared fixed point. The red line is the critical line that connects the two fixed points.

\vspace{0.5cm}

\newpage
\section{The gradient flow in scalar field theory}

To improve readability we will speak only of $\phi^4$ theory in this section. The $\phi^6$ term does not contain additional information relevant for the following discussion. The Euclidean action for $\phi^4$ theory is
\begin{equation}
  S = \int d^Dx\; \left( \frac12(\partial_{\mu}\phi)^2 + \frac12m^2\phi^2 + \frac{\lambda}{4!}\phi^4 \right).
\end{equation}
We define the flow equation \cite{Monahan:2015lha}
\begin{equation}
  \partial_t\bar{\varphi}_t(x) = \partial^2 \bar{\varphi}_t(x), \qquad \left.\bar{\varphi}_t(x)\right\rvert_{t=0} = \phi(x),
\label{eq:floweq}
\end{equation}
where the flow time $t$ and the flow field $\bar{\varphi}_t$ were introduced. The flow equation determines the evolution of $\bar{\varphi}_t$ along $t$. The flow field is bounded to be the scalar field $\phi$ at zero flow time. The solution to the flow equation (\ref{eq:floweq}) shows that the flow has a smoothing effect on the fields at the boundary which are smeared with radius $r=\sqrt{8t}$.

\begin{figure}
\centering
\begin{tabular}{c l l}
  \raisebox{-.2\height}{\includegraphics[width=3.5cm,trim={3cm 23.5cm 10cm 3.5cm},clip]{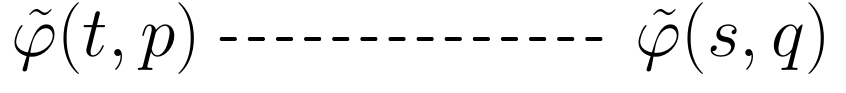}} &$\qquad$& ${\displaystyle \frac{1}{p^2+m^2}} \e^{-(t+s)p^2}$\\[.5cm]
  \raisebox{-.2\height}{\includegraphics[width=3.5cm,trim={3cm 23.5cm 10cm 3.4cm},clip]{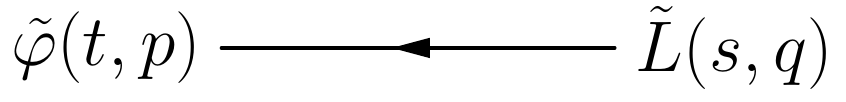}} &$\qquad$& $\theta(t-s) \e^{-(t-s)p^2}$\\[.5cm]
  \raisebox{-.4\height}{\includegraphics[width=2.5cm,trim={4.5cm 22.5cm 12cm 2.5cm},clip]{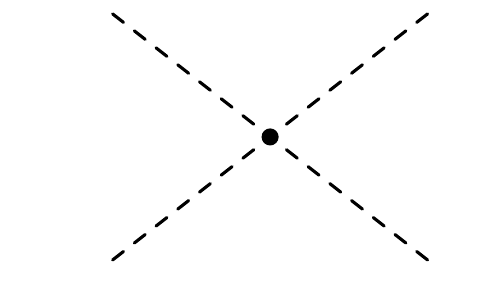}} &$\qquad$& $-\lambda$
\end{tabular}
\caption{Feynman rules in 3+1 dimensions for scalar $\phi^4$ theory plus gradient flow.}
\label{fig:FR}
\end{figure}

It is now possible to formulate a higher dimensional theory by considering the flow time as an additional direction \cite{Luscher:2011bx}. The action is then composed of the action of the boundary theory plus a part that accounts for the gradient flow which is implemented using a Lagrange multiplier field $L$,
\begin{align}
  &S = S_{\text{boundary}}+S_{\text{bulk}},\\
  &S_{\text{bulk}} = \int_0^{\infty}dt \int d^Dx\; L(t,x)\left(\partial_t-\partial^2\right)\varphi(t,x).
\end{align}
Integrating out $L$ in the path integral gives a delta function which ensures that the field $\varphi$ is the solution of the flow equation $\bar{\varphi}_t$. The Feynman rules of the $3+1$ dimensional theory are drawn in fig. \ref{fig:FR}. The flow field propagator includes the propagator of the scalar field at $\,t=s=0$. In addition, there is a propagator coming from the newly introduced field $L$. The $4$-point vertex exists only at the boundary since there is no additional interaction term in the bulk action.

The bulk fields $\varphi$ and $L$ do not require renormalisation. Since there is no interaction term in the bulk, there are no divergences coming from loop diagrams in the bulk. On the other hand, renormalising the bulk fields leads to infinite counterterms in the bulk for which no counter part in the form of loop divergences exists.

\begin{figure}
\centering
\begin{tabular}{c l l}
  \raisebox{-.4\height}{\includegraphics[width=2cm,trim={4.5cm 22.5cm 12cm 2.5cm},clip]{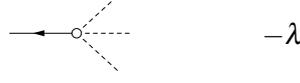}} &$\qquad$& $-\lambda$
\end{tabular}
\caption{Additional vertex to the Feynman rules when defining the flow as gradient of the action.}
\label{fig:FR2}
\end{figure}

If we were to choose the flow equation differently, e.g. as the gradient of the action as it is done for pure gauge theory \cite{Luscher:2010iy}, the flow equation reads
\begin{equation}
  \partial_t \bar{\varphi}_t = \partial^2\bar{\varphi}_t-m^2\bar{\varphi}_t-\frac{\lambda}{3!}\bar{\varphi}_t^{3}.
\end{equation}
The Feynman rules in the 3+1 dimensional theory change accordingly. There is now a mass in the exponential function in both propagators, and an additional vertex in the bulk, see fig. \ref{fig:FR2}.

Following \cite{Luscher:2011bx} one can compute the divergences of the self-energy of the flow field propagator in 4 dimensions. The divergences disappear in 3 dimensions. However, it turns out that the existence of the terms proportional to the couplings $m^2$ and $\lambda$ in the bulk action lead to non-renormalisable divergences. Renormalising the couplings $m^2=Z_m \;m_R^2$ and $\lambda=Z_\lambda\,\lambda_R$ yields for the modified bulk action
\begin{equation}
 S_{\text{bulk}} = \int_0^{\infty}dt \int d^Dx\; \left( L(t,x) \left(\partial_t-\partial^2+Z_m\;m_R^2\right)\varphi(t,x) + L(t,x)\;Z_\lambda\, \frac{\lambda_R}{3!}\;\varphi(t,x)^3 \right).
\end{equation}
The $Z$ factors contain divergences that enter into counterterms in the bulk for which no complementary loop divergences exists. Hence, it is not possible to have any bare couplings in the bulk action and we are left with the simple flow equation defined in eq. \ref{eq:floweq}.

\section{Lattice setup}

The naive discretisation of the complete action and the flow equation yields,
\begin{align}
  &\hat{S} = a^3\sum_n \left( \frac12(\hat{\partial}_{\mu}\phi)^2 + \frac12m^2\phi^2 + \frac{\lambda}{4!}\phi^4 + \frac{\eta}{6!}\phi^6 \right)\\[0.2cm]
  &\qquad \partial_t\,\varphi_t = \hat{\partial}^2\, \varphi_t, \qquad\qquad \left.\varphi_t\,\right\rvert_{t=0} = \phi,
\end{align}
where $\hat{\partial}$ is a lattice derivative, and $\hat{\partial}^2$ is the lattice Laplacian. The flow equation can be implemented by numerical integration, or one can use its solution directly.

\begin{figure}
\centering
\begin{tabular}{r c}
 \raisebox{8\height}{$\scriptstyle m^2$} & \includegraphics[width=6cm,trim={6cm 4cm 2cm 4.5cm},clip]{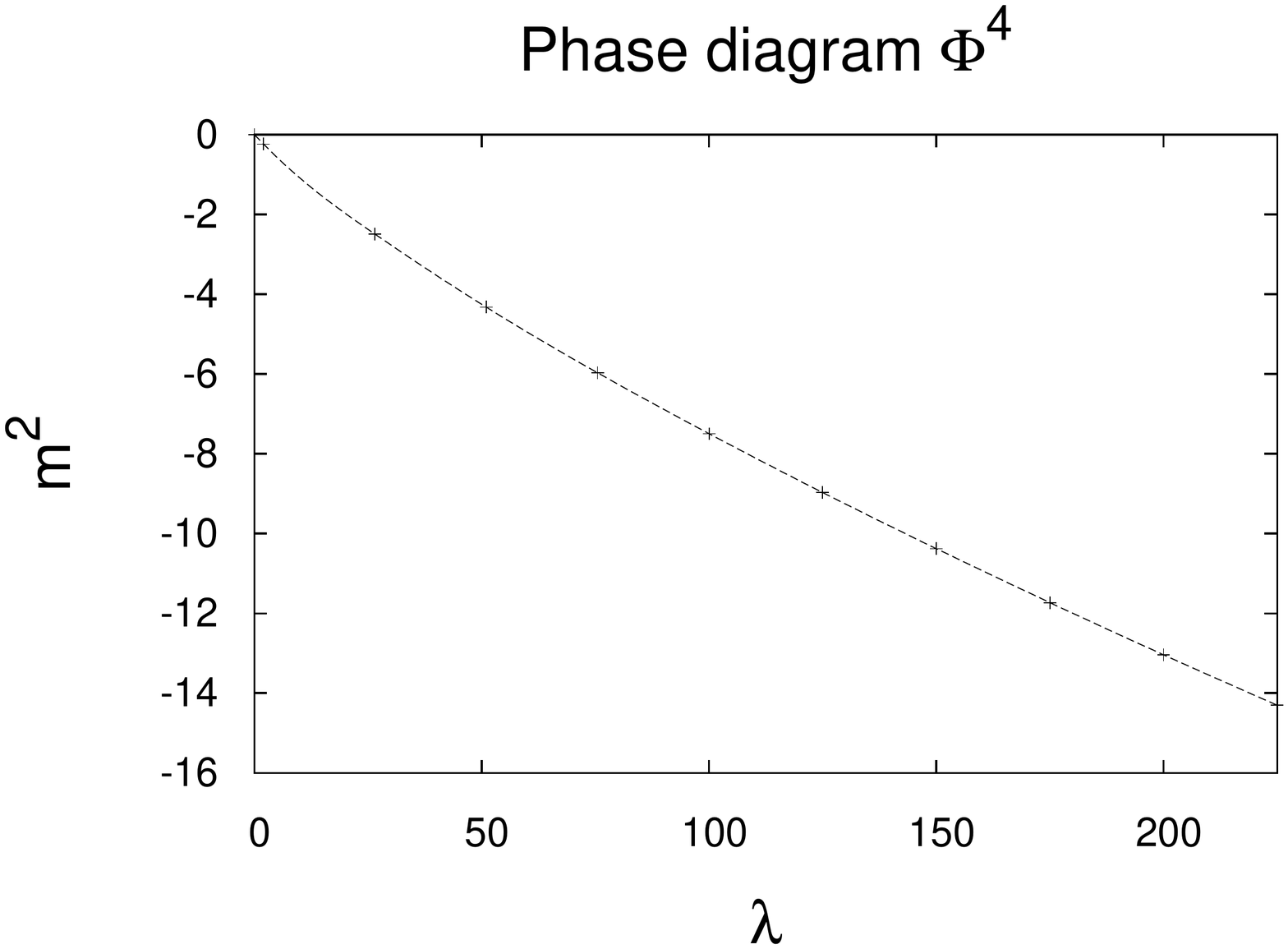}\\
 & \hspace{0.5cm} \raisebox{0.4\height}{$\scriptstyle \lambda$}
\end{tabular}
\caption{Phase diagram of $\phi^6$ theory at $\eta=0$.}
\label{fig:phasediag}
\end{figure}

The theory has two phases if we allow for negative bare mass square. Fig. \ref{fig:phasediag} shows the plot of the phase transition where the renormalised mass $m_R^2=0$, at $\eta=0$.

The algorithm used consists of two parts as suggested in \cite{Brower:1989mt}: a Swendsen-Wang update and a Metropolis update. Our measurements are sufficiently uncorrelated. The integrated autocorrelation time $\tau$ \cite{Wolff:2003sm} is monitored for all observables and is always about one half. It does not depend on the lattice size, the value of the flow time, or in which phase we simulate. The integrated autocorrelation times for different values of $\lambda$, $m$ and $L$ close to the phase transition for the observables action $S$ and flow field squared $\varphi^2$ are listed in tab. \ref{tab:autocorr}. Using a cluster algorithm ensures that we do not observe any critical slowing down in our simulations. Measurements are taken every 200 Metropolis and 40 cluster sweeps. The simulation time is short with e.g. about 8 hours for 33000 measurements on an $8^3$ lattice run on one core of an Intel Xeon E5620 at 2.4 GHz.
\begin{figure}
\centering
\begin{tabular}{c | c | c | c | c}
    $\lambda$& $m$ & $L$ & $\tau$(S) & $\tau(\varphi^2)$ \\\hline
    2 & 0.5231 & 8  & 0.5000 & 0.6989 \\
    2 & 0.4842 & 20 & 0.5507 & 0.5000 \\    
    100 & 2.7522 &  8 & 0.5000 & 0.5308 \\
    100 & 2.7449 & 20 & 0.5005 & 0.6244 \\
    175 & 3.4118 & 8  & 0.5049 & 0.5000 \\
    175 & 3.4239 & 20 & 0.5000 & 0.5437
\end{tabular}
\caption{Integrated autocorrelation times $\tau$ for different values of $\lambda$, $m$ and $L$ for action $S$ and $\varphi^2$.}
\label{tab:autocorr}
\end{figure}

\section{Ward identities and energy-momentum tensor}

In the continuum the translation Ward identity for scalar field theory reads
\begin{equation}
  \langle\, \delta_{x,\rho}P \,\rangle = -\langle\, P\;\partial_{\mu}T_{\mu\rho}(x) \,\rangle,
\end{equation}
where $P$ is an arbitrary probe observable, and $\delta_{x,\rho}$ is a local operator of translation defined by
\begin{equation}
  \delta_{x,\rho}P = \frac{\delta P}{\delta\phi(x)}\,\partial_{\rho}\phi(x).
\end{equation}
On the lattice on the other hand, the lattice regularisation breaks translation symmetry explicitly and an additional term appears in the Ward identity which now reads
\begin{equation}
  \langle\, \hat{\delta}_{x,\rho} \hat{P} \,\rangle = - \langle\, \hat{P} \left(\hat{\partial}_{\mu}\hat{T}_{\mu\rho} + \hat{R}_{\rho} \right)\,\rangle.
\end{equation}
The operator $\hat{R}_{\rho}$ that accounts for the explicit symmetry breaking vanishes in the $a\rightarrow0$ limit. However, subleading terms in $\hat{R}_{\rho}$ can combine with divergences and give finite contributions. Thus, $\hat{R}_{\rho}$ and the energy-momentum tensor require renormalisation. The renormalised translation Ward identity on the lattice is
\begin{equation}
  \langle\, Z_{\delta} \;\hat{\delta}_{x,\rho} \hat{P} \,\rangle = - \langle\, \hat{P} \left(\hat{\partial}_{\mu}[\hat{T}_{\mu\rho}] + [\hat{R}_{\rho}] \right)\,\rangle,
\label{eq:renTWI}
\end{equation}
with the renormalised $[\hat{T}_{\mu\rho}]$ following from the mixing of $\hat{R}_{\rho}$ with all terms that are of equal or lower dimension and that possess the same symmetry properties,
\begin{align}
  [\hat{T}_{\mu\rho}] &= \frac{c_1}{2}\;  \hat{\partial}_{\mu}\phi \hat{\partial}_{\rho}\phi + \delta_{\mu\rho} \Bigg(  \frac{c_2}{2}\; \phi^2 +  \frac{c_3}{2}\; \hat{\partial}_{\mu}\phi \hat{\partial}_{\mu}\phi +  \frac{c_4}{4!}\; \phi^4 +  \frac{c_5}{2}\; \sum_{\lambda} \hat{\partial}_{\lambda}\phi\hat{\partial}_{\lambda}\phi +  \frac{c_6}{6!}\; \phi^6 \nonumber\\
  &\quad +  c_7\; \phi \hat{\partial}^2 \phi +  c_8\; \phi\hat{\partial}_{\mu}\hat{\partial}_{\mu}\phi + c_9\Bigg) + c_{10}\; \left( \hat{\partial}_{\mu}\hat{\partial}_{\rho}-\delta_{\mu\rho}\hat{\partial}^2 \right) \phi^2.
\label{eq:renEMT}
\end{align}
Here, the Einstein summation convention does not hold and the only sum is given explicitly. The operator corresponding to $c_{10}$ can be used instead of $\phi \hat{\partial}_{\mu}\hat{\partial}_{\rho}\phi$ \cite{CALLAN197042}.

It is now necessary to determine the 10 coefficients $c_1$ to $c_{10}$, and $Z_{\delta}$ which can be computed independently \cite{DelDebbio:2013zaa}. Choosing a probe observable $\hat{P}_t$ that is a function of fields at non-zero flow time the coefficients can be tuned such that the energy-momentum tensor is finite and $[\hat{R}_{\rho}]\rightarrow0$ \cite{DelDebbio:2013zaa}. Up to subleading corrections, and noticing that two terms drop out in the Ward identity, we find that we need to solve a system of 8 equations with 8 (or more) skilfully chosen probe observables,
\begin{equation}
 Z_{\delta} \;V^{(k)} = - \sum_i c_i \;M^{(k,i)}.
\end{equation}
Here, $V^{(k)}$ stands for the 8 expectation values on the left hand side of eq. (\ref{eq:renTWI}) for 8 different probes, and $M^{(k,i)}$ is the matrix whose rows are made up of the derivative of the remaining 8 operators in $[\hat{T}_{\mu\rho}]$. The dilatation Ward identity can be used in order to determine the two coefficients that cannot be found with the translation Ward identity.

As a preliminary exercise, we computed $Z_{\delta}$ for $m_R^2>0$ using \cite{DelDebbio:2013zaa}
\begin{equation}
  Z_{\delta} \;\langle\, \Phi_1(T,z) \,a^3\sum_{y\in D}\hat{\delta}_{y,\rho}\,\Phi_2(t,x) \,\rangle = \langle\, \Phi_1(T,z) \,\hat{\partial}_{\rho}\Phi_2(t,x) \,\rangle + \mathcal{O}\left(\e^{-\frac{r^2}{16t}}\right).
\end{equation}
The calculation has been performed in a Monte Carlo simulation, as well as in lattice perturbation theory. For $\Phi_1(T,z)=\varphi(t,z)$ and $\Phi_2(t,x)=\varphi(t,x)$ the right hand side of the perturbative formula is
\begin{equation}
 -\frac{i}{L^3} \sum_p \e^{ip(z-x)} \frac{\hat{p}_{\rho}}{m^2+\hat{p}^2}\; \e^{-2t\hat{p}^2-iap_{\rho}/2} \,\left( 1+\mathcal{O}(\lambda) \right)
\end{equation}
and the left hand side yields
\begin{equation}
 - Z_{\delta} \frac{ia^3}{L^3}  \sum_{y\in D} J(t,x;0,y) \sum_p \e^{ip(z-x)} \frac{\hat{p}_{\rho}}{m^2+\hat{p}^2}\; \e^{-t\hat{p}^2-iap_{\rho}/2} \,\left( 1+\mathcal{O}(\lambda) \right).
\end{equation}
The sum over $y$ is not carried out over the entire lattice but over a smaller domain $D$. $\hat{p}$ is the lattice momentum, and $J(t,x;s,y)$ is the Jacobian of the transformation $\bar{\varphi}_s(y)\rightarrow\bar{\varphi}_t(x)$ \cite{Luscher:2009eq},
\begin{equation}
 J(t,x;s,y)=\theta(t-s) \frac{1}{L^3} \sum_p \e^{-(t-s)\hat{k}^2} \e^{ik(x-y)}.
\end{equation}

\begin{figure}
\centering
\begin{tabular}{r c}
 \raisebox{11.5\height}{$\scriptstyle Z_\delta$} & \hspace{-0.2cm}\includegraphics[width=6cm,trim={2.2cm 2.5cm 3cm 3cm},clip]{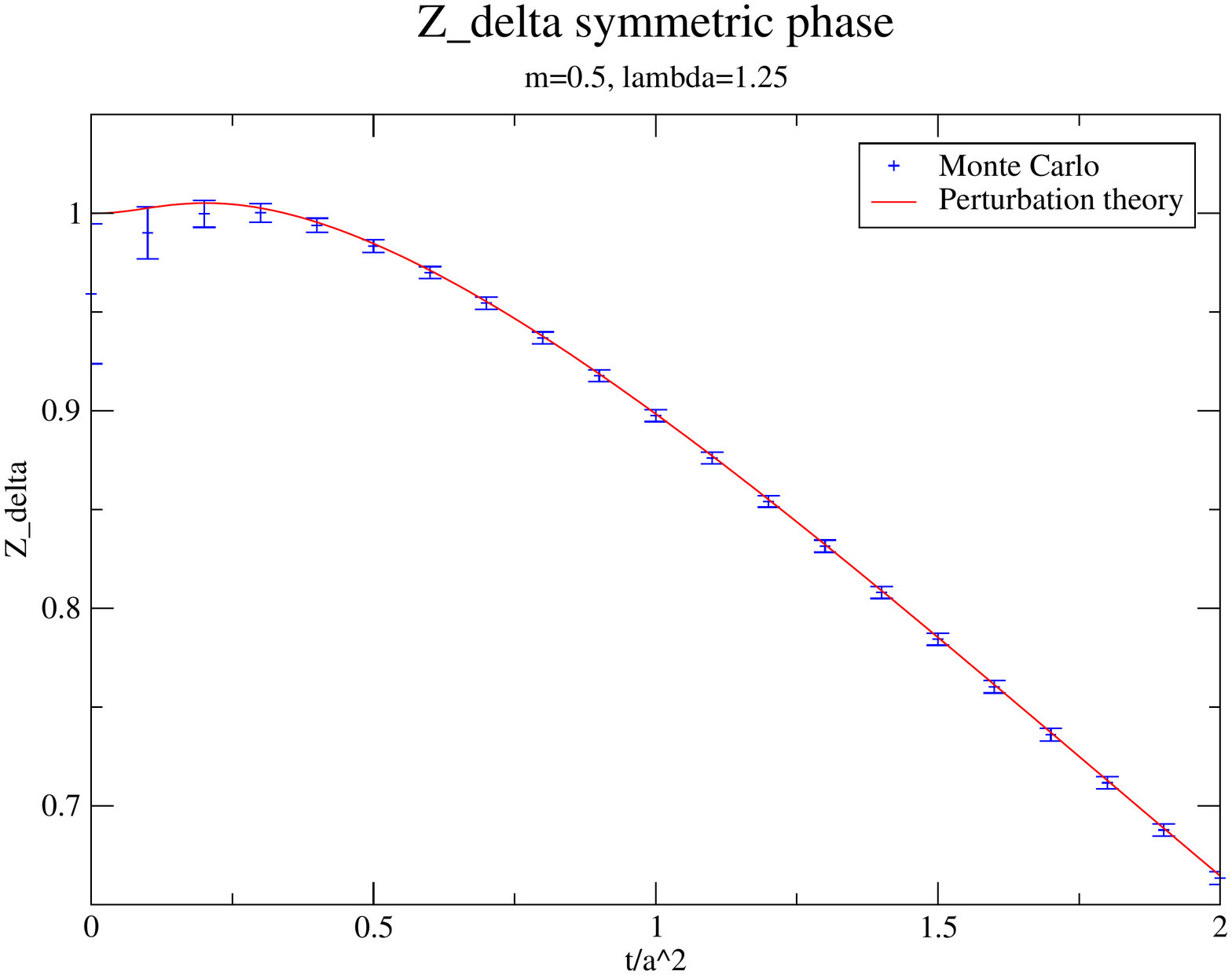}\\
 & \hspace{0.2cm}$\scriptstyle t/a^2$
\end{tabular}
\caption{Computation of $Z_\delta$ in a Monte Carlo simulation and lattice perturbation theory for $\lambda=1.25$.}
\label{fig:Zdelta}
\end{figure}

The numerical and analytical calculation agree. This can be seen in fig. \ref{fig:Zdelta} which shows the results for $Z_{\delta}$ of both methods as functions of the flow time at a random test value $\lambda=1.25$.

\section{Summary}

The gradient flow for scalar field theory in 3 dimensions, and the applicability of the Wilson flow for the determination of the renormalised energy-momentum tensor were discussed. We found that in principle it is possible to find a meaningful formulation of the energy-momentum tensor on the lattice. Inspecting a scalar field theory and a small number of dimensions has the advantage of being numerically cheap. At the same time we increase the number of operators that contribute to the renormalised energy-momentum tensor compared to gauge theory. The challenge is to make the method numerically effective. Future work will show if an explicit determination of the renormalised energy-momentum tensor is realisable in practice.

\section*{Acknowledgements}

We thank A. Patella and L. Keegan for fruitful discussions. The numerical computations have been carried out using resources from the HPCC Plymouth. LDD is supported by STFC, grant ST/L000458/1, and the Royal Society, Wolfson Research Merit Award, grant WM140078. SE is supported by the  Higgs Centre for Theoretical Physics. RP is supported by STFC (grant ST/L000458/1). AR is supported by the Leverhulme Trust (grant RPG-2014-118) and STFC (grant ST/L000350/1).

\end{document}